\documentstyle[aps,prl]{revtex} 
\draft 
 
\begin{document} 
 
\twocolumn[\hsize\textwidth\columnwidth\hsize\csname 
@twocolumnfalse\endcsname 
 
\title {Phenomenology of High Energy Neutrinos 
in Low-Scale Quantum Gravity Models} 
\author{J. Alvarez-Mu\~niz$^1$, F. Halzen$^2$, T. Han$^2$, D. Hooper$^2$} 
\address{ 
$^1$Bartol Research Institute, University of Delaware, Newark, DE 19716\\  
$^2$Department of Physics, University of Wisconsin, 1150 University Avenue,   
Madison, WI 53706 
} 
\date{June, 2001} 
 
\maketitle 
 
\begin{abstract} We show that neutrino telescopes, optimized for detecting   
neutrinos of TeV to PeV energy, can reveal threshold effects associated with   
TeV-scale gravity. The signature is an increase with energy of the cross   
section beyond what is predicted by the Standard Model. The advantage  
of the method is that the neutrino cross section is measured in an  
energy region where 
i)  the models are characteristically distinguishable and 
ii) the Standard Model neutrino cross section can be reliably   
calculated so that any deviation can be conclusively identified. 
\end{abstract} 
] 
 
Motivated by the absence of a self-consistent theory of quantum gravity and   
the unresolved hierarchy problem between the electroweak scale ($10^{2}$ 
GeV) and the Planck scale ($10^{19}$ GeV),  
a great deal of attention has been given to   
theories of low-scale quantum gravity which envision significant 
quantum gravity effects at an energy scale of 
the order of $M_s \sim$ 1 TeV \cite{ADD,RS}.   
In these scenarios, potentially large effects on high energy 
processes may occur due to the contributions from, {\it e.g.},  
Kaluza-Klein excitations of gravitons (KK) or other stringy states 
near $M_s$. An interesting motivation for models  
in which cross sections at TeV scale become  
enhanced is the ultra-high energy cosmic ray problem.  Protons above the GZK   
cutoff ($\sim 10^{19}$ eV) interact with the cosmic microwave background   
cataclysmically by the $\Delta$-resonance \cite{GZK,GZK2}. Thus, the cosmic   
ray events observed above this energy must be produced by local sources, or   
involve new physics.  Local sources of particles of such energy being    
unlikely, many exotic solutions have been proposed \cite{GZKsol}. A   
solution which has received a great deal of attention in recent literature   
proposes that neutrinos with enhanced cross sections at GZK energies 
constitute the highest energy cosmic rays \cite{uhecr3,uhecr4,vene}.   
This solution requires neutrino-nucleon cross   
sections on the scale of 10's of mbarnes.
Unfortunately, most scenarios of low-scale quantum gravity  
as low-energy effective theories are valid only up to 
the order of $\alt M_s$.  
Above this scale, the naive calculations typically violate 
unitarity \cite{uhecr3,uhecr4}.
One has to introduce some ad hoc unitarization scheme, since 
the fundamental theory, such as a realistic string theory, is  
yet unavailable. It is also very difficult to reliably predict  
the parton distribution functions needed at GZK   
energies in neutrino-nucleon interactions.  
For these reasons, studies of ultra-high energy ($\sim 10^{20}$ eV) 
quantum gravity enhancements to neutrino-nucleon interactions are 
extremely speculative. 
 
These problems are far more manageable at energies below or near  
$M_s$. Unitarity may not be violated at this scale,  
calculations are generally perturbative and the   
relevant parton distributions are known at these energies \cite{uhecr2}.   
The characteristics for different theoretical models can
be also qualitatively distinguishable near and slightly above
the threshold.
Therefore, the TeV regime provides a natural scale for  
probing the features of low-scale quantum gravity models.   
These tests include direct searches in colliders such as 
the Fermilab Tevatron and the CERN Large Hadron Collider 
(LHC) \cite{colliders,grav3,stringy}.   
This letter discusses another class of   
experiments capable of testing features of low-scale quantum gravity:   
multi-TeV to PeV neutrino astrophysics. 
 
We have recently witnessed first light of neutrino telescopes optimized to   
detect neutrinos in the TeV to PeV energy range \cite{amanda, baikal}. 
This is   
the range of laboratory energies where the onset of TeV-scale gravity  
effects on   
the neutrino cross section will first manifest itself. For a neutrino 
flux $\Phi$,   
the number of neutral currents events $N_{\nu}$ observed as hadronic 
showers in a neutrino detector of effective   
area $A$ is given by the convolution over energy of the quantity 
$A^{3/2} \times \Phi   
\times n \times \sigma_{\nu}$. Here $n$ is the density of the   
target that interacts with a neutrino with cross section $\sigma_{\nu}$ 
to produce a hadronic shower. In our discussion the detected neutrino  
flux plays a secondary role.  
It may represent the atmospheric neutrino flux or the  
flux of neutrinos of hundreds of TeV anticipated from gamma ray bursts.

An increase with energy of the cross section for neutrinos to 
interact with matter beyond a level calculated in the   
Standard Model will signal the onset of new physics including the 
increase anticipated as a consequence of TeV-scale gravity   
effects. It is important to recognize that the Standard Model cross 
section is   
computed from nucleon structure functions probed by HERA experiments in this   
energy range. The Standard Model baseline against which to measure  
new physics is known.  The important observation here is that 
due to the geometry of the Earth and detector, the angular 
distributions of events with new physics can be significantly 
different from the Standard Model prediction. These measurements, 
though challenging for existing instruments like AMANDA and the 
detectors in the Mediterranean, should be feasible 
for second-generation detectors such as IceCube. 
Unlike first generation neutrino telescopes, 
IceCube can separate interesting   
high energy events from the large background of lower energy atmospheric   
neutrinos by energy measurement. The instrument can identify high energy   
neutrinos over $4\pi$ solid angle, and not just in the lower hemisphere 
where they are identified by their penetration of the Earth, 
as is the case with AMANDA. Furthermore, due to the enhanced neutrino 
cross section from new physics, the Earth eventually 
becomes opaque to those high energy 
neutrinos. This leads to the striking phenomena that 
only down-going events from new physics at high energies are enhanced.

For the sake of illustration, we will examine three classes of  
low-scale quantum gravity/stringy models: 
\begin{itemize} 
\item[1).] ADD scenario \cite{ADD}: Large extra dimensions  
($R<0.1$ mm) with a flat Minkowski metric.  
The large effects are due to the high degeneracy of the  
light KK gravitons of mass $m^{}_{KK}\sim 1/R$. The theory saturates 
perturbative unitarity at $\sqrt s\agt M_s$ and some unitarization 
scheme has to be introduced. 
\item[2).] RS scenario \cite{RS}: One anti-de Sitter dimension with a 
non-factorizable ``warped'' geometry.  
The physical consequences relevant to our interests are the  
contributions from the KK gravitons of a TeV mass \cite{grav3}.  
\item[3).] Veneziano amplitude: Due to the lack of a fundamental 
theory of quantum gravity, a reasonable parameterization to the 
new physics at the scale $M_s$ perhaps is to include a sum of 
possible ``stringy'' states \cite{stringy}. The Veneziano 
amplitude serves for this purpose in describing the stringy physics,
before non-perturbative quantum gravity effects become overall dominant.
It also manifestly preserves unitarity \cite{vene}. 
\end{itemize} 
 
There are several free parameters in these models.  
The choices for these parameters were selected to  
illustrate a variety of phenomenological features and are not inclusive. 
First of all, we take the quantum gravity or string scale $M_s$ as
1 TeV throughout our studies. In addition, the ADD scenario  
is subject to the number of extra dimensions, although  
the effect under consideration is not sensitive to it \cite{colliders}.   
The RS model varies with the scale of the theory 
($\Lambda=e^{-kR}M_{pl}$) and the mass of graviton resonances ($m_g$).
In our analysis, we will only keep the first graviton resonance and 
ignore the heavier states. Models calculated with Veneziano amplitudes  
are characterized by two constants $a$ and $b$, which parameterize  
the Chan-Paton traces for string models \cite{vene}, in addition to $M_s$. 

Perturbative calculations for the ADD 
scenario lead to a parton-level cross section 
$\sigma(\nu g\to\nu g)\sim s^3/M_s^8$, which violates
partial wave unitarity $|Re(a_\ell)|<1/2$ at energies above 1.5$M_s$.  
To have a sensible estimate, partial wave amplitudes are taken
to saturate unitarity when the unitarity bound is reached.
Calculations are then made with only $S$-wave, with 5 leading
partial waves, as well as with all partial waves that need to
be unitarized. This variety of choices reflects our ignorance 
of how nature chooses to restore unitarity at a higher energy 
scale. However, we consider the $S$-wave unitarization a 
conservative estimation for the ADD scenario.
The RS model is less likely to violate unitary as the  
rapid growth of amplitudes only occurs way above graviton mass  
scale. The Veneziano amplitudes automatically respect 
unitarity bounds. 
 
\begin{figure}[thb] 
\vbox{\kern2.4in\special{ 
psfile='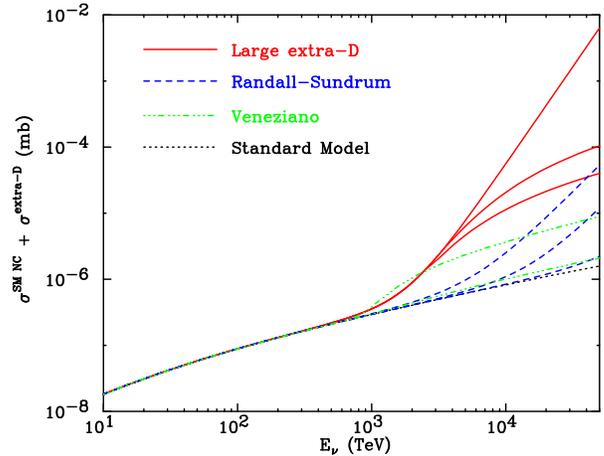' 
angle=0 
voffset=-78
hscale=45 
vscale=45}} 
\caption{Neutrino-nucleon cross sections in a variety of models 
compared to the Standard Model neutral current prediction.  
ADD (large extra dimension) models are
for all, 5 and 1 partial waves, 
up to unitarity saturation (top to bottom).  
RS models are shown for $\Lambda$=1 TeV, $m_g$=500 GeV; $\Lambda$=1 TeV, 
$m_g$=1 TeV; $\Lambda$=3 TeV, $m_g$=500 GeV (top to bottom).  
Models using Veneziano amplitudes are for $a=b=5$ and $a=b=0$ 
(top to bottom). $M_s$=1 TeV for all models.}
\label{one} 
\end{figure} 

Figure \ref{one} depicts the sum of the neutral current neutrino 
nucleon cross section within the Standard Model  
and the corresponding cross section due to new physics
for the different models discussed above. 
For the reasons explained earlier, we wish to explore new physics
not too far from the threshold, and thus do not address the energy 
range above $E_\nu=5\times 10^4$ TeV. 
The increase in the cross section beyond the Standard Model  
due to TeV-scale quantum gravity starts at a threshold energy 
$E_{\rm s}^\nu \sim 10^3$ TeV, corresponding to a neutrino-nucleon
center-of-mass energy near 1 TeV. Indeed, the three models under
consideration have distinctive characteristics near the threshold,
as clearly seen in Fig.~\ref{one}.
For the large extra dimension
scenario (ADD), the three (solid) curves correspond to
those of unitarized amplitudes ($S$-wave only, 5 partial waves,
and all partial waves that saturate unitarity). Up to an energy
$E_\nu=5\times 10^3$ TeV, the three solid curves are identical, 
indicating the $S$-wave dominance. At higher energies, there
will be more partial waves to reach the unitarity bound. 
We consider the $S$-wave unitarization scheme a 
conservative representation for this model. We see that
above $E_{\rm s}^\nu$ the effects of the new physics 
in ADD should be clearly visible for which the increase in the  
cross section beyond the Standard Model expectations could be
orders of magnitude at $E_\nu=5\times 10^4$ TeV. 
For the RS scenario, the influence with a single
KK graviton exchange should be also visible if the graviton
is not too heavy. The result with
$\Lambda=1$ TeV and a graviton mass $m_g=500$ GeV could
lead to a maximum increase in the cross section of a factor 
of 30 at $E_\nu=5\times 10^4$ TeV. On the other hand, the
effects would be unobservable if we take $m_g=1$ TeV and 
$\Lambda=3$ TeV. 
The stringy model of Veneziano amplitude \cite{vene} 
predicts an increase about a
factor of 6 with $a=b=5$ at $E_\nu=5\times 10^4$ TeV,
while the enhancement is negligibly small if we take $a=b=0$.

\begin{figure}[thb] 
\vbox{\kern2.4in\special{ 
psfile='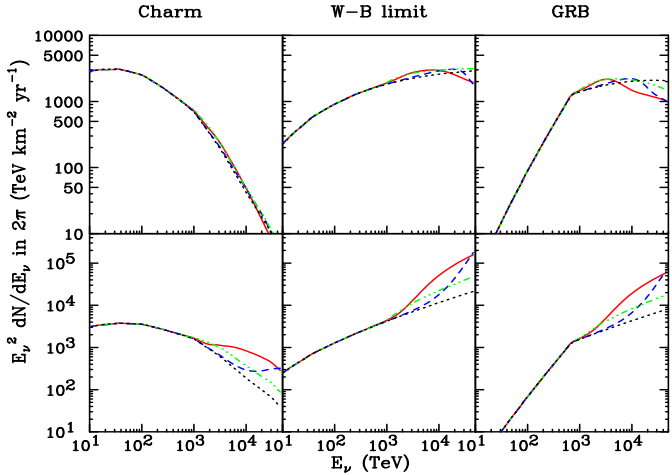' 
angle=0 
hoffset=-20 
voffset=-85
hscale=50 
vscale=45}} 
\caption{Energy distribution of $\nu_\mu+\bar\nu_\mu+\nu_e+\bar\nu_e$ 
induced shower events in IceCube. 
The upper panels show the up-going neutrino events and the lower panels the  
down-going events. The panels are labeled with the corresponding  
theoretically predicted neutrino flux (see text).  
In each panel, the solid line is the event rate 
for $\sigma^{\rm SM}~+~\sigma^{\rm ADD}$(S-wave);  
the dashed 3-dotted for 
$\sigma^{\rm SM}~+~\sigma^{\rm Veneziano~}$ with $a=b=5$; 
the dashed line for
$\sigma^{\rm SM}~+~\sigma^{\rm RS}$ 
with $m_{\rm g}=500$ GeV and  
$\Lambda=1$ TeV and, the dotted for $\sigma^{\rm SM}$ 
alone.  $M_s$=1 TeV for all models.}
\label{two} 
\end{figure} 
 
Figures 2 and 3 illustrate the qualitative behavior explained above
in the IceCube detector.
We explore three different theoretical predictions of the neutrino 
flux. These have been chosen mostly for illustrative purposes.  
The panel labeled charm refers to maximal predictions of  
neutrinos from the decay of charmed particles produced by  
cosmic ray interactions in the atmosphere \cite{Costa:2000jw}.  
The W-B refers to the Waxman and Bahcall (W-B) limit on the 
neutrino flux from  
astrophysical sources that are optically thin to proton-photon  
and proton-proton interactions. This represents a flux of 
$E_\nu^2~\Phi_\nu=2\times 10^{-8}~{\rm GeV~(cm^2~s~sr)}^{-1}$ 
\cite{Waxman:1999yy}. GRB labels the neutrino flux  
from Gamma Ray Bursts, energetic explosions in the Universe which occur 
at a rate of $\sim 1000$ per year. The flux accounts for 
fluctuations in the distance to individual GRBs and in the energy  
they release in the form of gamma 
rays \cite{Halzen:1999xc,Alvarez-Muniz:2000st}.  
Note that the unusual features in  
the angular distribution of GRB neutrinos are a result of these 
fluctuations \cite{Halzen:1999xc}.  Although these results can 
widely vary, those shown are typical. 

\begin{figure}[thb] 
\vbox{\kern2.4in\includegraphics{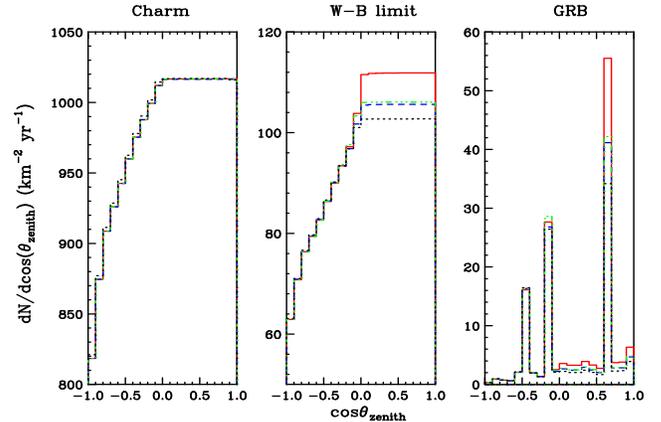}} 
\caption{$\cos(\theta_{\rm zenith})$ distribution of 
$\nu_\mu+\bar\nu_\mu+\nu_e+\bar\nu_e$ 
induced shower events in IceCube. $\cos(\theta_{\rm zenith})=-1$ 
corresponds to vertical up-going neutrinos, 
and $\cos(\theta_{\rm zenith})=0$ 
to neutrinos coming from the horizon. The panels are labeled  
with the corresponding theoretically predicted neutrino flux 
that has been used to obtain the event rate (see text). 
The legend for different models is the same as in Fig.~2.}  
\label{three} 
\end{figure} 

The upper panels of Fig.~2 
show the up-going neutrino induced shower events 
($-1<\cos\theta_{\rm zenith}<0$)
and the energy spectrum for down-going shower events  
($0<\cos\theta_{\rm zenith}<1$) is shown in the lower panels. 
The event rate peaks at the neutrino energies  
at which the product  
$\Phi(E_\nu) \times \sigma_\nu(E_\nu)$ 
maximizes. 
Figure 3 shows the zenith angle distribution of the 
neutrino induced shower events for the same three neutrino fluxes.
Notice that the shower rate
has a contribution from Standard Model charged
current neutrino interactions which is present in all plots. 
The angular distribution of the up-going events is dominated
by absorption in the Earth whereas for down-going events
absorption is not important and the angular distribution
reflects the detector's effective volume to contained showers.

Little sensitivity to new physics is anticipated 
when looking for up-going events, while increase in the event 
rate for down-going due to new physics may be visible if the 
neutrino fluxes are large enough at high energies. 
Within the large extra dimension model (ADD), the signal is
observable even for the conservative scenario with
the $S$-wave unitarization. 
Reasonable sensitivity is expected to the RS model with
a single graviton exchange if $m_{\rm g} \alt 500$ GeV,
and similar sensitivity is predicted for the Veneziano scenario
if $a,b\sim 5$. The excess of  
events due to new physics in the charm case is very 
small ($<1$ per ${\rm km^2}$ and year) because the flux 
drops quickly in the energy range where the  
cross section increases, although there is large uncertainty 
for the atmospheric neutrino flux at this energy regime.
The W-B flux gives a sizeable signal due to the larger 
neutrino flux rate near the threshold energies of new physics.
The integrated numbers of events are summarized in Table I
for the different neutrino fluxes and new physics scenarios studied. 
The numbers in parenthesis are the event rates with muons.
They include 
contributions from muons in charged current $\nu_\mu+\bar\nu_\mu$
interactions
within the Standard Model and muons produced in hadronic showers
induced by $\nu_\mu+\bar\nu_\mu+\nu_e+\bar\nu_e$.
The IceCube detector is sensitive
to down-going neutrino induced muons above 1 PeV since it can 
determine the energy
of the events and hence separate them from the background
produced by cosmic rays in the atmosphere, which is smaller than
the neutrino induced muon rate for energies above $\sim 1-10$
PeV \cite{hundertmark}.

\begin{figure}[thb] 
\vbox{\kern2.4in\includegraphics{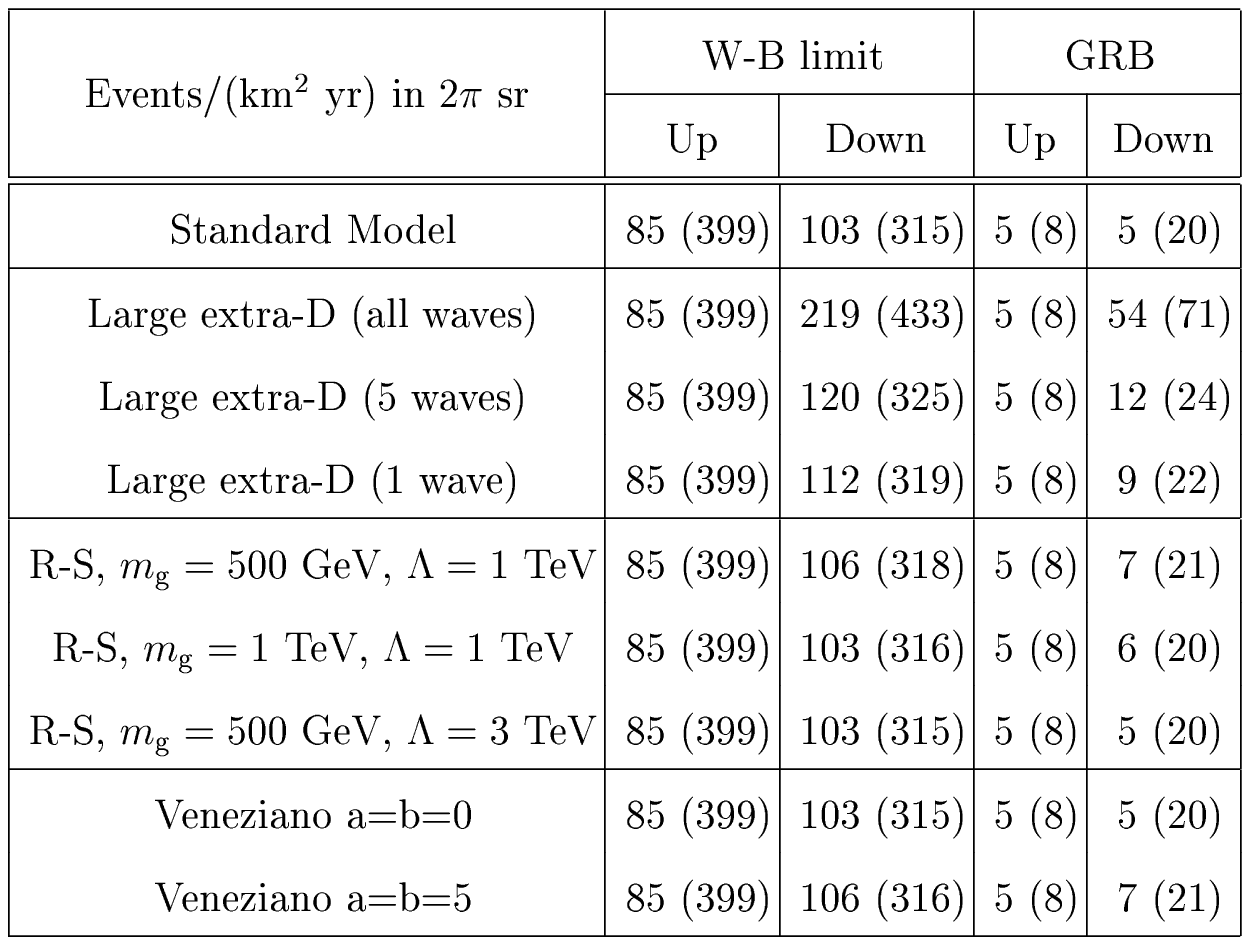}} 
\caption{
$\nu_\mu+\bar\nu_\mu+\nu_e+\bar\nu_e$
shower event rate and $\mu$ event rate (in parenthesis) per ${\rm km^2~yr}$ 
in IceCube for different theoretically predicted $\nu$ fluxes and 
$\nu$-nucleon cross sections. $\mu$ events include
contributions from muons in charged current $\nu_\mu+\bar\nu_\mu$ 
interactions
within the Standard Model and muons produced in hadronic showers
induced by $\nu_\mu+\bar\nu_\mu+\nu_e+\bar\nu_e$.  
The muon energy threshold is 
$E_\mu$=100 GeV, the shower energy threshold is 1 TeV and the maximum
neutrino energy is $5\times 10^4$ TeV.}
\label{four} 
\end{figure} 

In summary, TeV scale quantum gravity or stringy models
have distinctive characteristics near the new physics 
threshold and may be probed in large high energy neutrino 
telescopes.
The ADD scenario with TeV scale quantum gravity has
distinctive phenomenological features for high energy
neutrino scattering.
We illustrated this point by a conservative unitarized $S$-wave
approximation. 
The RS model with a single graviton exchange
can also provide observable signatures
if the graviton mass is lighter than about 1 TeV.
TeV Stringy models using Veneziano amplitudes can produce 
interesting features in the case with large Chan-Paton
trace factors. Studies of these features may allow for discovery 
of such models, or for stronger  
constraints on the scale of new physics. Neutrino astrophysics 
can provide a means to complement searches for TeV 
scale quantum gravity in collider  
experiments. 
 
{\it Acknowledgments}: This work was supported in part by a DOE grant No.  
DE-FG02-95ER40896  
and in part by the Wisconsin Alumni Research Foundation. 
J.A-M is supported  
by the NASA grant NAG5-7009.

\end{document}